\documentclass[twocolumn]{aastex63}

\usepackage{xcolor}
\usepackage{ulem}

\begin{document}
\title{Symmetries of CMB Temperature Correlation at Large Angular Separations}

\correspondingauthor{Craig Hogan}
\email{craighogan@uchicago.edu}

\author[0000-0002-7500-6576]{Ray Hagimoto}
\affiliation{University of Texas at San Antonio}
\affiliation{University of Chicago, 5640 South Ellis Ave., Chicago, IL 60637}
\author[0000-0002-1433-8841]{Craig Hogan}
\affiliation{University of Chicago, 5640 South Ellis Ave., Chicago, IL 60637}
\affiliation{Fermi National Acclerator Laboratory, Batavia, IL 60510}
\author[0000-0002-8671-1190]{Collin Lewin}
\affiliation{Steward Observatory, University of Arizona, 933 North Cherry Avenue, Tucson, AZ 85721}
\affiliation{University of Chicago, 5640 South Ellis Ave., Chicago, IL 60637}
\author[0000-0003-3315-4332]{Stephan S. Meyer}
\affiliation{University of Chicago, 5640 South Ellis Ave., Chicago, IL 60637}
\submitjournal{\apjl}

\begin{abstract}
A new analysis  is presented of the angular correlation function $C(\Theta)$ of  cosmic microwave background (CMB) temperature at large angular separation, based on published maps derived  from  {\sl WMAP} and {\sl Planck} satellite data, using different models of astrophysical foregrounds. It is found that using a common analysis, the results from the two satellites are very similar. In particular, it is found that  previously published differences  between  measured values of $C(\Theta)$ near  $\Theta=90^\circ$ arise mainly from different choices of masks in  regions of largest Galactic emissions, and that demonstrated measurement biases are reduced  by eliminating masks altogether.  Maps from both satellites are shown to agree with  $C(90^\circ)=0$ to within  estimated statistical and systematic errors, consistent with an exact symmetry  predicted in a new holographic quantum model of  inflation.
\end{abstract}



\section{Introduction}
In the standard cosmological  model,  initial conditions are set by a combination of a uniform inflationary background space-time, and perturbations from fluctuations of a quantum field vacuum matched to linearized gravity. For an appropriate choice of background parameters, the model leads to a perturbation power spectrum in good agreement with measurements of cosmic large scale structure, and with the angular spectrum of anisotropy in the cosmic microwave background radiation (CMB) \citep{2013ApJS..208...20B,Akrami:2018vks}.


Recently, a new class of models has been proposed for the quantum mechanics of inflationary initial conditions \citep{Banks:2018ypk,PhysRevD.99.063531,Hogan:2019rsn}.  In these holographic or ``spooky'' models, the quantum system is based not on fields, but on coherent states of space-time  structure.  Quantum field  states collapse coherently on comoving spatial hypersurfaces, but holographic quantum-geometrical  states collapse coherently on the  inflationary horizon --- the inbound null cone that arrives at an observer at the end of inflation.  Nonlocal entanglement  leads to emergent classical scalar curvature perturbations with new  correlations in direction, and over a range of comoving scales.

The new approach is motivated by holographic, emergent theories of quantum gravity, in which
space, time, gravity, and perhaps locality itself emerge statistically from a holographic quantum system based on null surfaces, such as light cones and horizons \citep{Jacobson1995,Padmanabhan:2013nxa,Jacobson:2015hqa}. 
Coherent quantum horizons have been extensively studied in the contexts of 
black holes \citep{Hooft:2016itl,Hooft:2016cpw,Hooft2018,Solodukhin:2011gn} and  anti-de Sitter spaces \citep{Ryu:2006bv,Ryu:2006ef,Natsuume:2014sfa}. 
Quantum fluctuations of the emergent null surfaces in such theories can be  much larger than the Planck-length variance predicted by  standard linearized gravity, based on effective field theory; their effects on macroscopic scales might even produce detectable effects in laboratory experiments \citep{Hogan:2015b,Hogan:2016,Verlinde:2019xfb}.

In holographic inflation,   coherent quantum-geometrical fluctuations of the  horizon are the main source of cosmic  scalar curvature perturbations \citep{PhysRevD.99.063531}.   
The scaling of emergent scalar curvature perturbations  produces the same nearly-scale-invariant, slightly tilted power spectrum  as the standard scenario, so it duplicates every result of standard cosmology that only depends on the power spectrum. The model leads to  specific predictions:  an  inflationary expansion rate in Planck units  given approximately by the observed scalar perturbation amplitude $Ht_P\approx A_S \approx 2\times 10^{-9}$; a similar value for the tensor-to-scalar ratio $r$;   an inflationary potential with a derivative given by approximately the inverse Planck mass; and a small or vanishing  intrinsic dipole and  global mean curvature.  The most conspicuous observable signature is that  nonlocal phase correlations of primordial curvature lead to a distinctive pattern of relic cosmic large scale structure and  CMB anisotropy, the subject of this work.

\subsection{Predicted pattern from holographic inflation}

The holographic absence of one independent degree of freedom  creates new  symmetries in directional relationships of the primordial potential around any observer,  associated with causal constraints on primordial information \citep{Hogan:2019rsn}. The new symmetries are predicted to appear today as precise constraints on angular correlations that limit the range of  possible patterns of CMB anisotropy. In contrast, the ensemble of possible skies in the standard quantum inflation model, based on independent random fluctuations in modes of a quantum field vacuum, includes many realizations incompatible with holographic symmetries.   
The holographic symmetries may enable a unified interpretation of  some features in the pattern of large angle CMB anisotropy  long known to disagree with expectations in the standard scenario\citep{WMAPanomalies,Schwarz:2015cma,Ade:2015hxq,Akrami:2019bkn}. 

In particular, emergent holographic correlations lead to new symmetries of the CMB temperature two-point correlation function, 
\begin{displaymath}
C(\Theta) \equiv \langle{\delta T}_a {\delta T}_b \rangle_{\angle ab = \Theta},
\end{displaymath}
the all-sky average of the product $\delta T_a \delta T_b$ of  CMB temperature  deviations from the overall mean of a dipole-subtracted map, for all pairs of points $a,b$  at angular separation $\angle ab = \Theta$.  

The most robust  prediction of  holographic inflation is that {\it $C(\Theta)$  exactly vanishes at  $\Theta=90^\circ$}, which follows simply from the  independence of primordial perturbations along axes in orthogonal directions. As discussed below,  this symmetry is {\sl not} a property of the standard scenario. 
The prediction  $C_{90} \equiv C(90^\circ) = 0$ for primordial curvature should be preserved in the observed sky temperature anisotropy:  it  is not affected by the magnitude or direction of any unmeasured intrinsic cosmic dipole, or
by  post-inflation effects that modify the distribution of temperature  on small angular scales from the pattern of  primordial scalar curvature;  for example, it is not changed by integrated Sachs-Wolfe anisotropy, or by  Doppler contributions to temperature anisotropy.

A similar causal symmetry may also lead to  a nearly-vanishing correlation at $\Theta=30^\circ$, although in this case a primordial directional symmetry of curvature is not exactly preserved by temperature anisotropy.  It is also possible that the quantum states of the inflationary horizon display an antipodal antisymmetry similar to black hole horizons, in which case it should   generically produce significant fine-grained anticorrelation  at  angles approaching $180^\circ$.

\subsection{Previous analysis}

While  $C(\Theta)$ is well known to have small values at large angular separation \citep{Bennett_2003,WMAPanomalies,Ade:2015hxq,Akrami:2019bkn}, there has not been particularly close scrutiny of its exact value, both because little particular significance  is attributed to them in standard cosmology, and because structure  is  contaminated by  astrophysical emission correlated on large angular scales.  A more accurate analysis of $C(\Theta)$  is needed to test predictions of an exact or nearly-exact symmetry. 


\begin{figure}[tpb]
\begin{centering}
\includegraphics[trim=0pt .2in .3in .9in,width=0.9\linewidth]{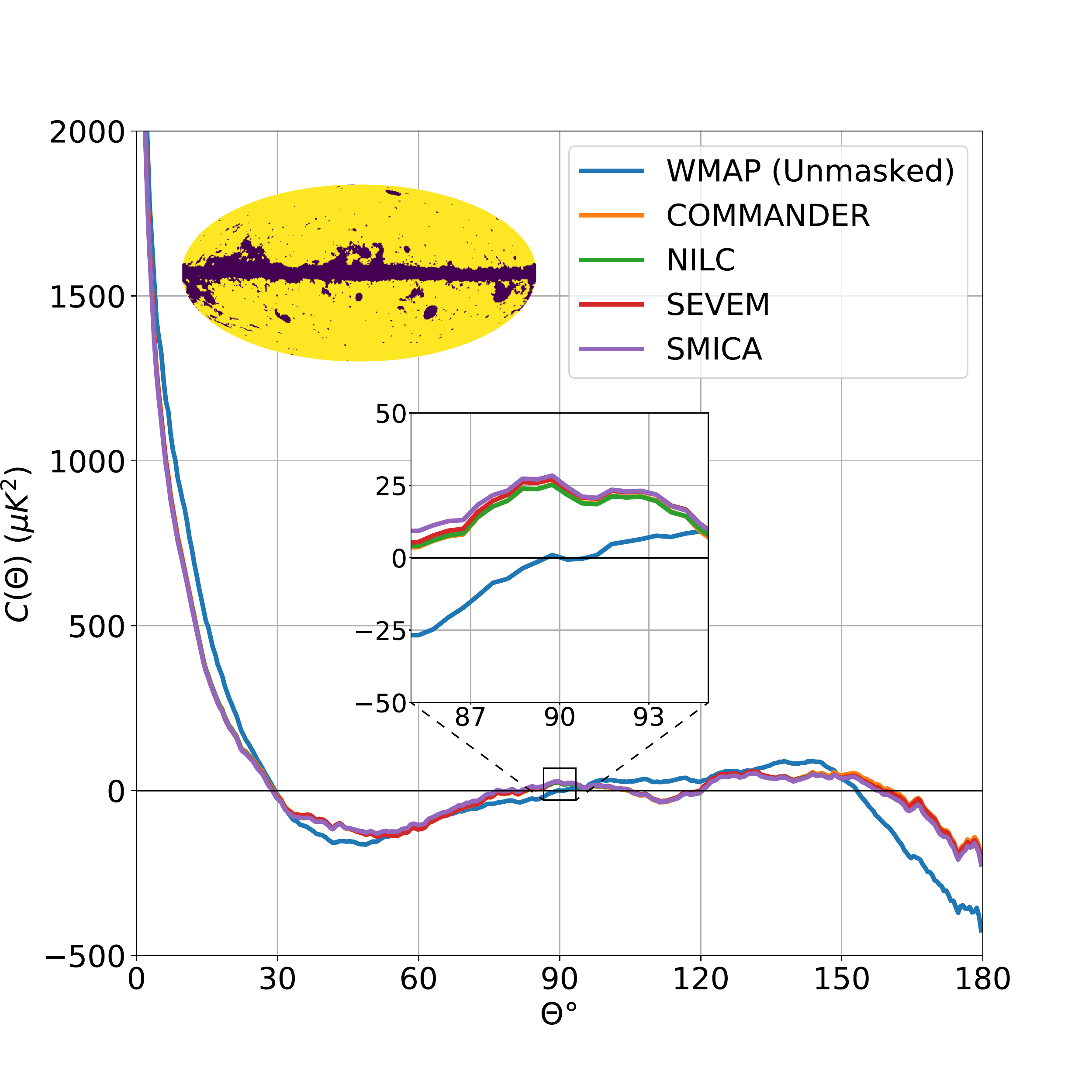}
\par\end{centering}
\protect\caption{Comparison of {\sl Planck} and {\sl WMAP} $C(\Theta)$ functions. The lines show our recalculation of $C(\Theta)$. To make these plots, the {\sl Planck} maps were  made omitting regions of sky  in the ``strong'' mask shown in the upper left, and the {\sl WMAP}  map was unmasked. These plots approximately reproduce the respective published {\sl Planck} and {\sl WMAP} correlation functions \citep{WMAPanomalies,Ade:2015hxq}. The inset shows a blow-up of the region centered at $\Theta = 90^\circ$, showing an apparent disagreement between {\sl Planck} and {\sl WMAP} correlation functions, and a good agreement of all four {\sl Planck} pipelines.
We have traced the bulk of  this disagreement to the use of  different foreground masks.  When all the foreground-reduced maps from {\sl Planck} and {sl WMAP} are analyzed with the same masks, the differences in the values of $C(\Theta)$ are very much reduced. \label{Planck-WMAP-compare}}
\end{figure}


The  {\sl  WMAP} team and the {\sl Planck} collaboration have both published papers on the correlation function and related large angular scale statistics. 
The most detailed study of the {\sl  WMAP}  correlation function, based on 7 years of data, includes a comparison with a standard-inflation ensemble prediction \citep{WMAPanomalies}. 
For {\sl Planck}, temperature anisotropy and statistics for the full data set are analyzed in  \cite{Ade:2015hxq};  
Fig.~2 in that  paper 
 shows  the correlation function for the  four {\sl Planck} pipelines, made with a common mask.
 

In plots of $C(\Theta)$ in \cite{Ade:2015hxq}, all of the {\sl Planck}  maps appear to  agree with each other, but on close examination they disagree in detail with the published  {\sl WMAP} plot found in \cite{WMAPanomalies}.
We reproduce the previous results with our reanalysis of masked maps as displayed  in Fig.~\ref{Planck-WMAP-compare}. We were led to undertake the present study by the prediction $C_{90}=0$ of holographic inflation, which appears to be inconsistent with the {\sl Planck} measurement.

In this Letter, we describe a new uniform analysis of these maps  to make an improved estimate of the  cosmological correlation, particularly at $\Theta = 90^\circ$ and $>160^\circ$.
 We present a  comparison of $C(\Theta)$  for all the maps with uniform pixelation, smoothing, binning and  masking.

\section{Data}

Our analysis is based on public data releases of foreground-corrected maps of the CMB temperature from the {\sl  WMAP} and {\sl Planck} satellites. These are the highest quality data available with consistent and homogeneous full sky coverage. 

In the case of {\sl  WMAP}, we use  the nine-year,
Internal Linear Combination (ILC) map. This map is a reconstruction of the thermal CMB sky, separated from  foreground components using multiband frequency information  \citep{2013ApJS..208...20B}.

In the case of {\sl Planck}, there are four separate  pipelines, using different foreground models, priors, and statistical methods:  Commander, NILC, SEVEM, and SMICA.
Initial results were described  by  \cite{PlanckXII};
both data and methods were later updated, including the full mission dataset for temperature anisotropy
\citep{Planckmaps}.  
The updates of the four maps \citep{Planckforegrounds} are  adopted here.

The foreground models use a broad range of complementary assumptions and statistical approaches,  with different biases. The analyses are mostly independent, but there are caveats: for example,   the Commander model used some external data at low frequencies, including ground-based 408 MHz data and {\sl WMAP} itself, so it is not entirely independent of {\sl WMAP}; at the same time, the model  was created independently of the {\sl WMAP} ILC, and uses new {\sl Planck} data over a wider range of frequencies.
Regarding the four {\sl Planck} maps, we quote from \citep{Planckmaps}:  ``we generally consider Commander to be the preferred solution on large and intermediate angular scales
\dots [and] we confirm our preference for the SMICA map for analyses that require full-resolution observations in temperature.''  



\section{Method}
To calculate $C(\Theta)$ we used the python wrapper for the Hierarchical Equal Area isoLatitude Pixelization ({\sl HEALPix}) scheme \citep{Healpix}, to read in the appropriate FITS CMB map (either {\sl WMAP}, or one of the four {\sl Planck} pipelines) downgraded to a resolution defined by $N_{\text{side}}=128.$ We then directly computed the pairwise temperature products $\delta T_a \delta T_b$ and angular separations $\angle ab$, storing the values in an array so that each row contained ($\angle ab, \delta T_a \delta T_b$) as an element. Masking was accomplished by skipping over any point where $a$ or $b$ was in the mask. Next, we found all entries for angular separations in a bin size of $0.5^\circ$, taking the angular separation $\Theta$ as the mean of the $\angle ab$ in the bin, and the unweighted mean of the corresponding $\delta T_a \delta T_b$ as the value for $C(\Theta)$. 
As a check of pixelation errors,  $C(\Theta)$ was measured with each unmasked map for 20 random orientations. The resulting spread in all cases is less than $\pm 1\,\mu{\rm K}^2$. 

\begin{figure}[htbp]
\begin{centering}
\includegraphics[trim=0pt .2in .3in .9in,width=0.9\linewidth]{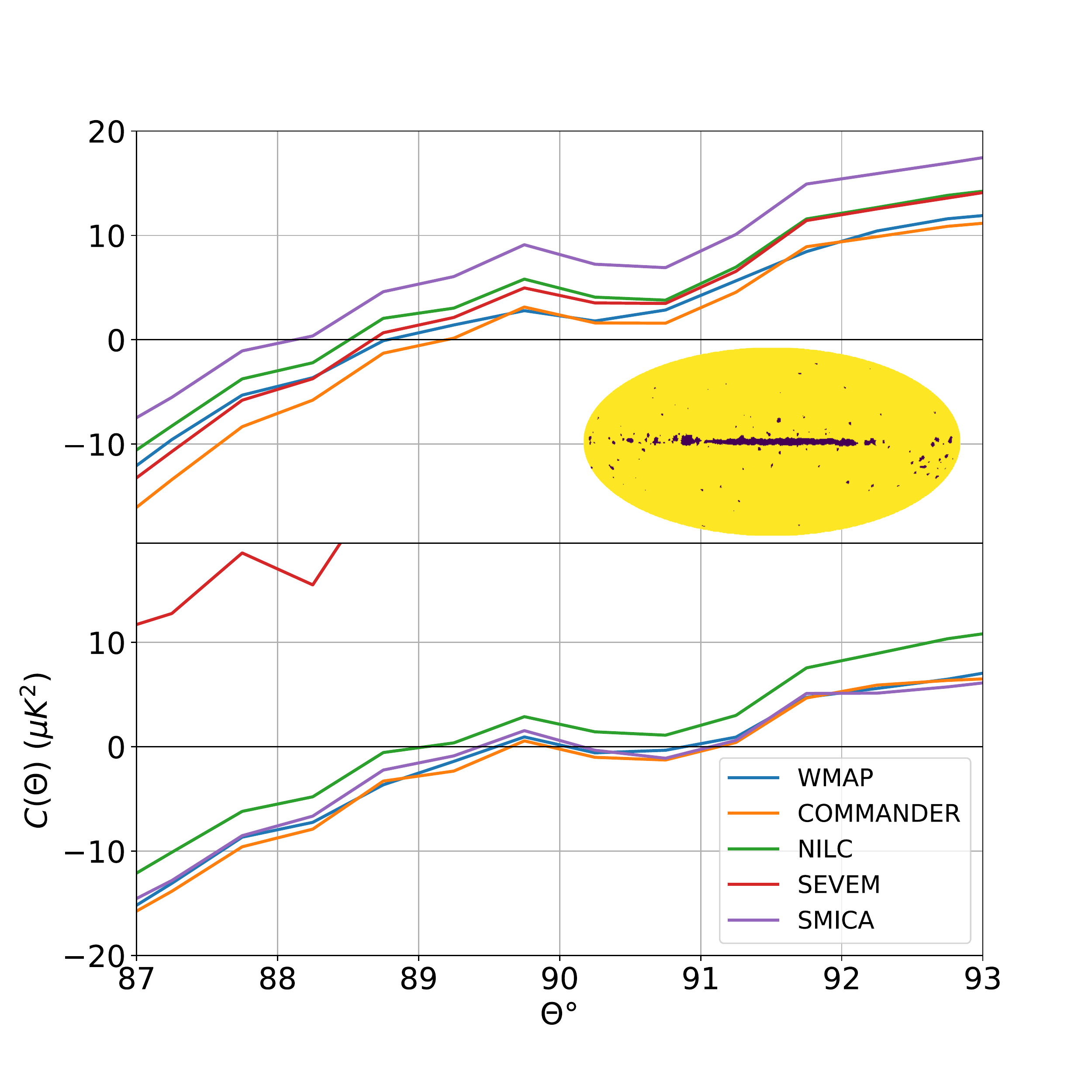}
\par\end{centering}
\protect\caption{ Estimate of the effect of masking near $\Theta=90^\circ$. The top panel shows the $C(\Theta)$ of the four {\sl Planck} maps and the {\sl WMAP} ILC map, made with a weak mask (as shown in the inset); the four {\sl Planck} maps now agree with {\sl WMAP}, and with each other. The bottom panel shows the 5 functions with no masking. With the exception of the SEVEM map, the agreement  is much tighter.  The overall  agreement between maps increases as the mask is reduced and then  eliminated. 
 \label{compare80100}}
\end{figure}

To investigate the effect of masking on the correlation function we chose a ``weak'' mask and a ``strong" mask from the second {\sl Planck} public release database\footnote{The masks are shown in as insets in Fig.~\ref{Planck-WMAP-compare} and Fig.~\ref{compare80100}. Filenames are \texttt{COM\_CMB\_IQU-nilc-field-Int\_2048\_R2.01\_full} and \texttt{COM\_CMB\_IQU-common-field-MaskInt\_2048\_R2.01} for the weak and strong masks respectively.}.  
Systematic  errors introduced by  foreground masks are estimated by randomly rotating the weak  and strong masks. 
In these tests, one map,  SEVEM,  appears as an outlier compared with the other maps. 
The simplest interpretation of why the SEVEM map appears as an outlier   is that it includes a larger residue of foreground contamination than the others. A visual inspection of the SEVEM map indeed shows much more emission aligned with features in the Galaxy than the other maps.


Excluding SEVEM,  at  $\Theta = 90^\circ$,  weak masking introduces a variation range of $\sim \pm 6.5 \mu {\rm K}^2$ and strong masking introduces a variation range of $\sim \pm 27 \mu {\rm K}^2$.
This range of variation provides an estimate of the effect on the measured value of $C_{90}$  typically produced in each map by masking alone. 

\bigskip

In Fig.~\ref{compare80100},
we show a comparison of the five maps with uniform weak masks and no masks near $90^\circ$. Consistent with the test just described, a weak mask as shown substantially reduces the variation from the comparison between {\sl WMAP} and {\sl Planck} shown in Fig.~\ref{Planck-WMAP-compare}. Not masking at all reduces the variation between the maps still further, for all of the maps except SEVEM. 
 It is important to note that the ``no mask" maps are based on  model restorations (or ``inpainting'') that interpolate over small regions at the Galactic center and inner Galactic plane. The small-scale additions, to the extent  that they are made independently for the different maps,  seem to influence $C(\Theta)$   very little near $90^\circ$. 
 
\begin{figure}[htbp]
\begin{centering}
\includegraphics[trim=0pt .2in .3in .5in,width=0.9\linewidth]{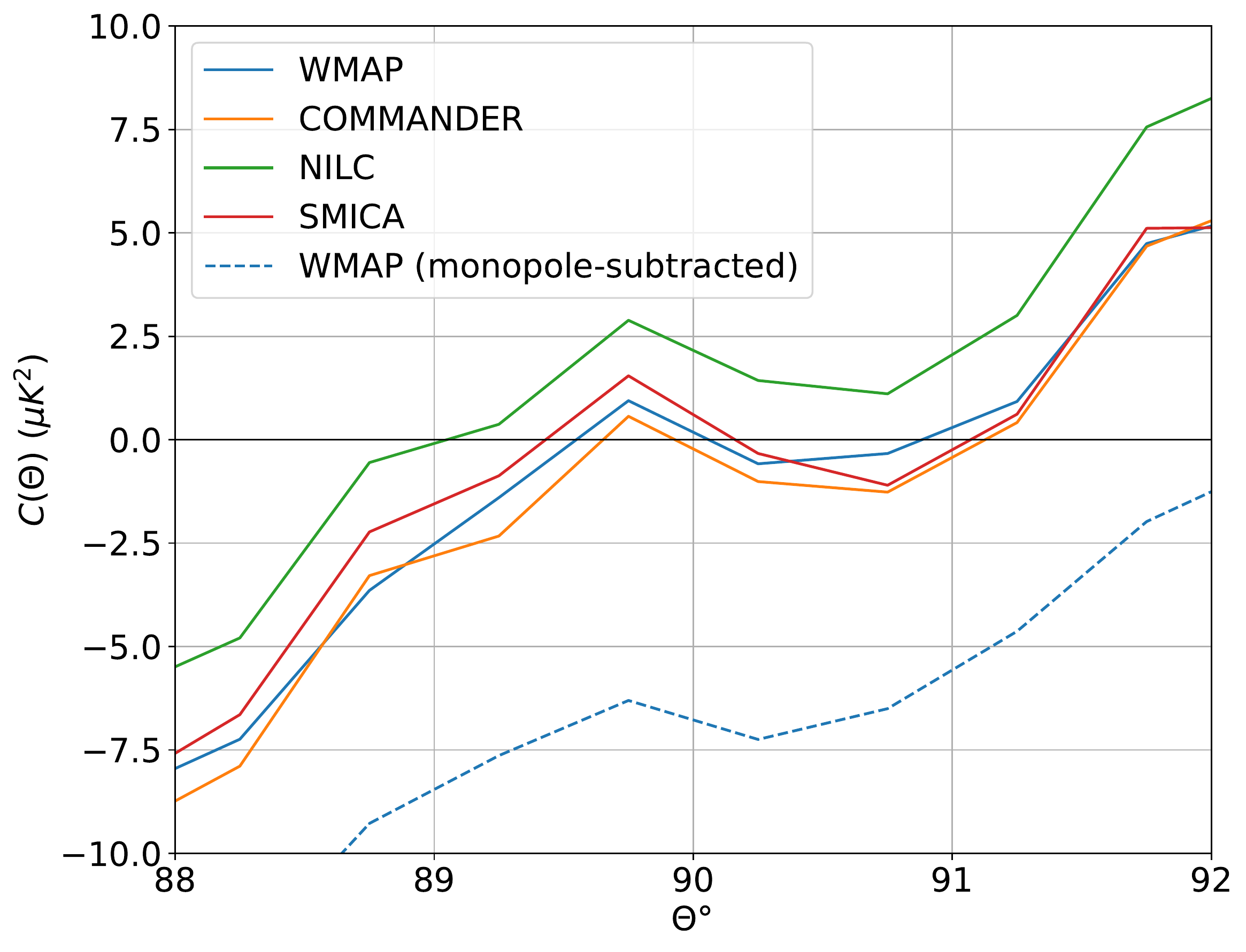}
\par\end{centering}
\protect\caption{A closer view of $C(\Theta)$ near $\Theta=90^\circ$ made with no masks.  SEVEM  lies outside the range of this plot. The  other {\sl Planck} maps span  a total range of less than $3\,\mu {\rm K}^2$. The agreement between the  {\sl WMAP},  SMICA and Commander maps  is exceptional, with  total range at $90^\circ$ of only 0.83 $\mu{\rm K}^2$. The dotted line shows $C(\Theta)$ for the monopole- and dipole-  subtracted {\sl WMAP} ILC map. 
\label{compare8793}}
\end{figure}

A closer view of  $C(\Theta)$ for the  foreground-subtracted, unmasked maps is shown in Fig.~\ref{compare8793}. 
At this level of scrutiny, an
 additional systematic uncertainty appears from  monopole and dipole harmonics present in the {\sl WMAP} ILC, which are much larger than  in the {\sl Planck} maps. 
The presence of a  monopole  is incompatible with  a  $\delta T$ map. As shown in Fig.~\ref{compare8793},  its removal reduces the value of $C_{90}$ by $\sim 6.7 \mu K^2$, outside the range  of the {\sl Planck} maps. 


The overall agreement between the different $C(\Theta)$'s  constrains both independent statistical and independent systematic errors in the measurements, although any in-common systematic errors (such as  in-common inaccuracies in foreground models) could be larger.  We adopt the spread of all the curves as an estimate of the systematic error.


\section{Results}

An important result of this analysis is to demonstrate the remarkable agreement between $C(\Theta)$ calculated from the  {\sl WMAP} and {\sl Planck} maps of the CMB sky over a  large range of $\Theta$. The agreement is significantly  better than was apparent from previously published plots and is the result of a uniform analysis. 

\bigskip



\subsection{Correlation Function at 90 Degrees}

The results illustrated in  Figs.~\ref{compare80100} and \ref{compare8793} show that over a range of $\Theta$ near $90^\circ$, the differences in measured $C({\Theta})$ from instrument and scan strategy systematics and statistical errors in the  measurements are small compared to  errors introduced by independent foreground subtraction, which are in turn small compared to errors known to be introduced by  masking.

Our most striking new result is that there is general agreement on a very small absolute value of $C_{90}$ in  {\sl WMAP} and {\sl Planck } maps analyzed with no mask.   All of the maps show this result except one outlier, SEVEM, which can be excluded due to residual foreground contamination. The  range spanned by the  other three {\sl Planck} maps is   $-0.22 \mu {\rm K}^2 < C_{90} < +2.16 \mu {\rm K}^2$. 
Combining these with the monopole-subtracted {\sl WMAP}, the lower end of the range extends to $-6.7 \mu {\rm K}^2$.  Although it is possible that an in-common foreground subtraction error between the different schemes perfectly cancels a  nonzero primordial signal, the best overall estimate is consistent with a value of     $C_{90}$ very close to zero.

 


\begin{figure}[htbp]
\begin{centering}
\includegraphics[width=0.9\linewidth]{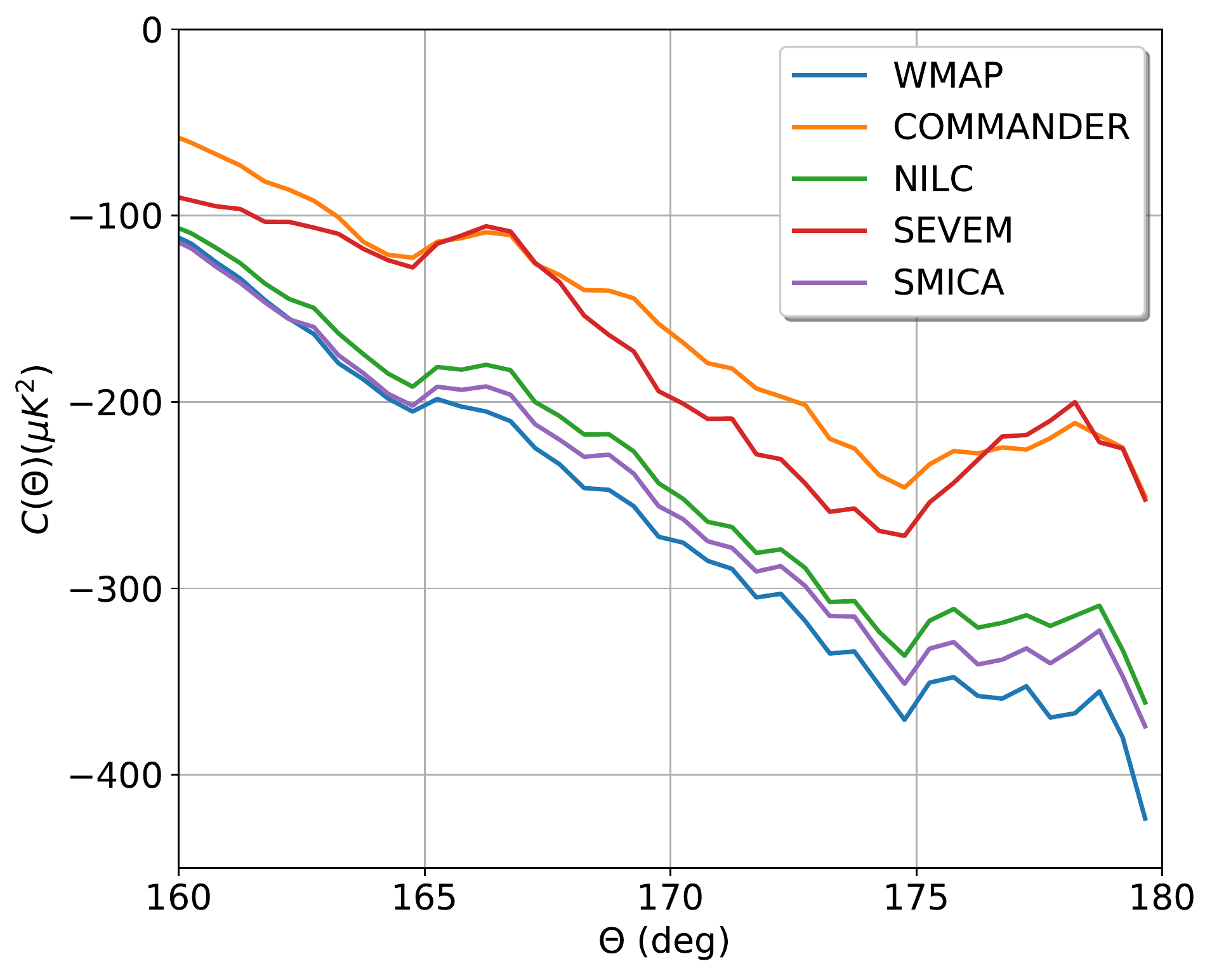}
\par\end{centering}
\protect\caption{Plots of $C(\Theta)$  for the five unmasked maps at $\Theta>160^\circ$.   \label{compare160180}}
\end{figure}

\subsection{Correlation Function Above 160 Degrees}

At large angles $\Theta> 160^\circ$,
the differences among the maps are much larger than at $90^\circ$.   All of the maps show a similar shape, a  significant negative correlation and a significant negative slope, but differ in  magnitude  by nearly a factor of two (Fig.~\ref{compare160180}). 
The differences in foreground modeling appear to have a larger effect than  systematic  bias  produced by masking, in contrast to the situation near $90^\circ$.  
Some but not all  of the difference can be accounted for by  contributions $\delta C_1(\Theta)\propto \cos(\Theta)$ due to unsubtracted dipole components, which vanish for some but not all of the maps.





\section{Interpretation}

\begin{figure}[htbp]
\begin{centering}
\includegraphics[width=0.9\linewidth]{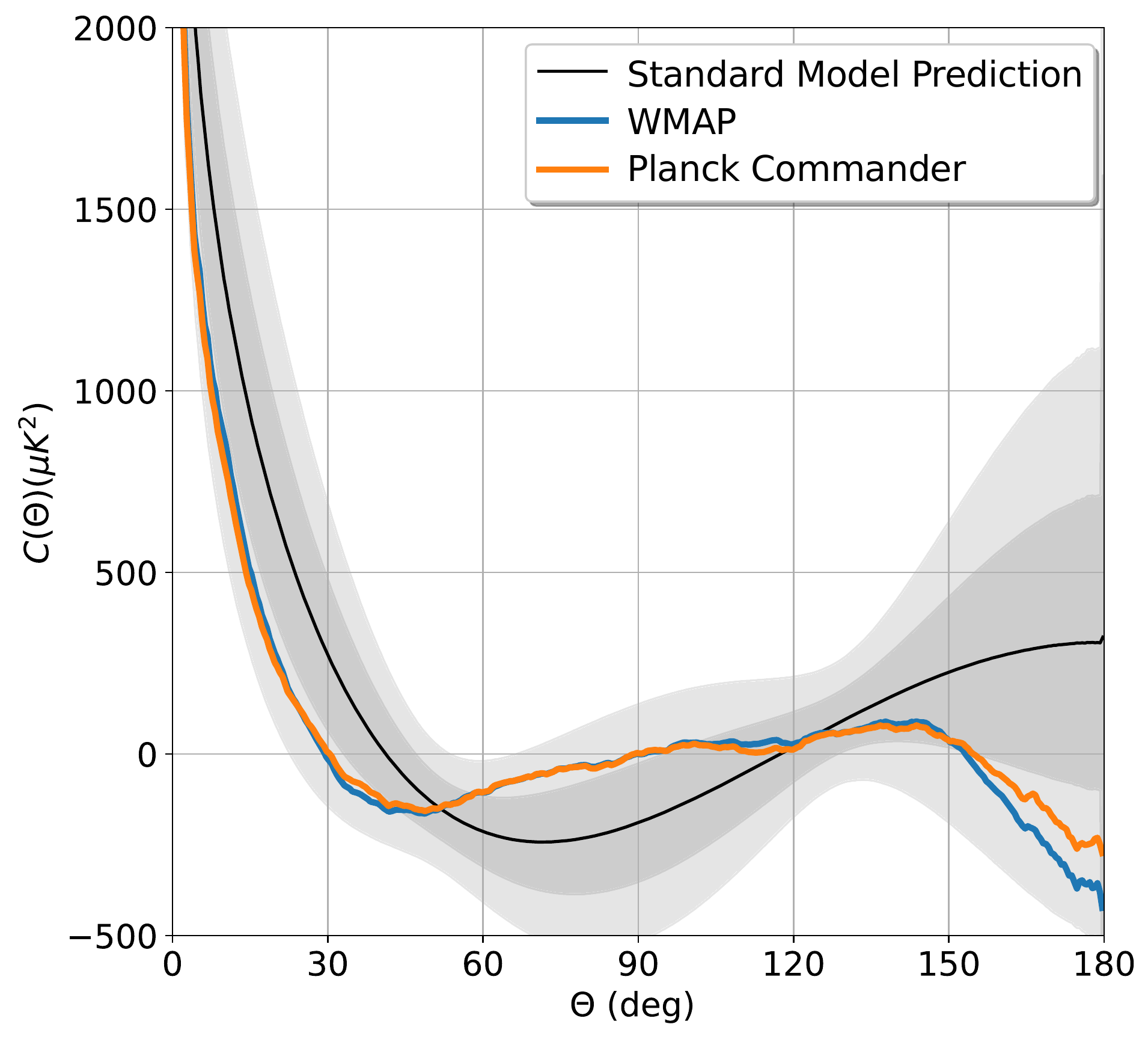}
\par\end{centering}
\protect\caption{Comparison of the unmasked {\sl WMAP} and {\sl Planck} Commander correlation functions. Also shown as a black line is the mean, the 65\% (dark band) and 95\% (lighter band) range of $C(\Theta)$ calculated from an ensemble of  sky maps made using   predictions of standard quantum inflation.
  \label{comparesimulations}}
\end{figure}

Standard inflation theory  predicts an ensemble of possible correlation functions. For the actual sky, which is just one realization, this leads to a  large range of possible values of $C(\Theta)$, due to the cosmic variance of fundamentally independent  modes.  The standard interpretation is that the specific measured value of $C(\Theta)$ has little particular significance, since it is just one realization.
Even so, as previously noted \citep{Bennett_2003,WMAPanomalies,Schwarz:2015cma,Ade:2015hxq,Akrami:2019bkn}
and as shown in Fig.~\ref{comparesimulations}, measured values of $C(\Theta)$  depart significantly from those expected in the standard picture.


Holographic inflation predicts  that there should be universal correlation properties in the angular domain.
In this model, some  specific features of large angle relationships, such as $C_{90}=0$, can be understood as predictable consequences of holographic symmetries of a quantum-geometrical horizon wave function, rather than as statistical flukes in a random ensemble \citep{Hogan:2019rsn}. 
This kind of global, nonlocal
connection between large-$\ell$ and large-$\Theta$ properties appears to violate statistical homogeneity and isotropy in the standard picture, but  in  holographic inflation, they arise from the emergence of the local rest frame and global  metric  from   quantum relationships with any observer based on covariant causal diamonds that have no preferred velocity or direction, and entangle structure on all scales and directions. Simply put, \emph{holographic causal  symmetries in the angular domain appear to be  miraculous in the natural harmonic basis of the standard scenario}.

The difference between these  scenarios becomes  significant if  theory is compared with specific, precisely determined values.
To illustrate this  point with the   current analysis, consider the   nearly-null value of $C(\Theta)$ at $90^\circ$ shown in Fig.~\ref{compare8793}, compared with   with the large range  $> 300 \mu{\rm K}^2$  of predicted values from standard quantum inflation, as shown in Fig.~\ref{comparesimulations}. 
In the standard picture, the close agreement with zero found in the sky at $90^\circ$ would be spoiled by 1$\sigma$ variations of  even a  few of the hundreds of harmonic power coefficients ($C_\ell$'s)  at the map resolution.
For example, 0.52\% of standard realizations  produce $C_{90}$ by chance in the  range spanned by SMICA,{\sl WMAP}, Commander and NILC ($-0.22 \mu {\rm K}^2 < C_{90} < +2.16 \mu {\rm K}^2$). A  larger fraction (1.5\%) falls within the larger range also encompassed by  the monopole-subtracted {\sl WMAP}. If an additional constraint is added at $\Theta=30^\circ$ using the range of measured values, $-20\mu {\rm K}^2 < C(30^\circ) < + 17 \mu {\rm K}^2$, only one out of 12,813 standard-model realizations agrees with both constraints. A holographic model could produce $C(\Theta)=0 $ at these angles by symmetry, consistent with both constraints.

Another, less precise example is the increasingly negative correlation found at separations $\Theta>160^\circ$, which confirms the  
odd-parity power asymmetry  previously found  via harmonic analysis \citep{Ade:2015hxq,Akrami:2019bkn} to be anomalous in the standard picture at the 0.2\% level for  $\ell$ up to about 30. Both results show that opposite points on the sky tend to have opposite values, even at a resolution of a few degrees.



\section{Conclusion}

Our analysis shows overall consistency between independent measurements by  different satellites, and among several  independently developed  foreground subtraction schemes. 
We show that better estimates  of  $C(\Theta)$ are likely possible with existing data, which may be used to  test new quantum models of inflation. 
We  defer a more detailed comparative likelihood analysis of quantum inflation models to future work that incorporates more detailed attention to effects of foreground models. 

In the future, improved  measurements of polarization,  especially  all-sky polarization maps with more  comprehensive spectral information,  should enable better separation of the primordial pattern of scalar curvature on the horizon from other cosmological and astrophysical sources of anisotropy.
It may also be  possible to measure holographic directional correlations of primordial curvature in the pattern of   3D large scale structure, with sufficiently large and complete galaxy surveys. 


\acknowledgments
 {This work was supported by the Fermi National Accelerator Laboratory (Fermilab), a U.S. Department of Energy, Office of Science, HEP User Facility, managed by Fermi Research Alliance, LLC (FRA), acting under Contract No. DE-AC02-07CH11359, and by a National Science Foundation REU program at the University of Chicago.  We acknowledge the use of HEALPix/healpy and the Legacy Archive for Microwave Background Data Analysis (LAMBDA), part of the High Energy Astrophysics Science Archive Center (HEASARC);   the NASA/ IPAC Infrared Science Archive, which is operated by the Jet Propulsion Laboratory, California Institute of Technology, under contract with the National Aeronautics and Space Administration; and  UTSA's SHAMU and UA's Ocelote HPC clusters. We are grateful to the referee for helpful comments.}

\bibliography{spookyCMB}
\bibliographystyle{aasjournal}
\end{document}